\documentclass[prb,preprint,superscriptaddress,groupedaddress,showpacs,preprintnumbers,amsmath,amssymb]{revtex4}
\usepackage{graphicx}
\usepackage{subfigure}

\begin{document}

\title{Shape effects on localized surface plasmon resonances in 
metallic nanoparticles}
\author{Titus Sandu}
\affiliation{National Institute for Research and Development in Microtechnologies,
126A, Erou Iancu Nicolae street, 077190, Bucharest, ROMANIA}

\date{\today}

\begin{abstract}
The effect of smooth shape changes of metallic nanoparticles on localized 
surface plasmon resonances is assessed with a boundary integral equation 
method. The boundary integral equation method allows compact expressions of 
nanoparticle polarizability which is expressed as an eigenmode sum of terms 
that depends on the eigenvalues and eigenfunctions of the integral operator 
associated to the boundary integral equation method. Shape variations change 
not only the eigenvalues but also their coupling weights to the 
electromagnetic field. Thus rather small changes in the shape may induce 
large variations of the coupling weights. It has been found that shape 
changes that bring volume variations greater than 12{\%} induce structural 
changes in the extinction spectrum of metallic nanoparticles. Also the 
largest variations in eigenvalues and their coupling weights are encountered 
by shape changes along the smallest cross-sections of nanoparticles. These 
results are useful as guiding rules in the process of designing plasmonic 
nanostrucrures. 
\end{abstract}

\maketitle
\section{Introduction}
Localized surface plasmon resonances (LSPRs) have their origin in the 
interaction of metallic nanoparticles (NPs) with light. The advancements 
made in the last decade have enabled the use of LSPRs for light manipulation 
at the nanoscale \cite{Halas2010}. The applications of LSPRs include enhanced 
sensing and biosensors \cite{Nie1997,Kabashin2009}, cancer 
imaging and therapy \cite{Loo2005}, plasmonic lasers \cite{Hill2007} 
and spasers \cite{Stockman2008}, enhanced nonlinearities, etc. Metallic NPs can 
be made by chemical synthesis or by lithographic techniques. Thus a large 
variety of shapes have been obtained by chemical synthesis. Beside the 
spheres, the shapes include nanorods \cite{Yu1997}, cubes \cite{El-Sayed2001,Sau2004}, triangular prisms \cite{Okada2004}, tetrahedra 
\cite{El-Sayed2001}, and hexagonal prisms \cite{Sau2004}. In contrast, in 
``top-down'' or lithographic techniques the shape of the metallic NPs are 
``flatter'' such that structures like disks \cite{Haynes2003}, dimers of 
disks \cite{Rechberger2003} or bowtie structures \cite{Fromm2004} have 
been successfully fabricated. 

Optical properties associated with LSPRs are determined by the shape, size, 
structure, and local dielectric environment of the NPs \cite{Kelly2003, 
Bohren1998}. Along with the electric field enhancement created 
around NPs, the effect of the dielectric surrounding the metallic NPs is 
widely used in sensing by monitoring the shift of LSPR absorption peaks with 
respect to the local dielectric changes. The absorption peak redshifts as 
the embedding refractive index is increased \cite{Sandu2011,Link1999}. On the other hand, in the quasistatic approximation, when 
the particle size is much smaller than the wavelength, the size of particle 
doesn't play any role \cite{Bohren1998}. Thus, in the quasistatic 
approximation, a metallic nanosphere exhibits just a single LSPR which is 
dipolar, irrespective of the size. However, as the radius is increased and 
the quasistatic approximation is no longer valid, the LSPR of the nanosphere 
shifts toward infrared and higher multipolar resonances emerge in the 
spectrum \cite{Kelly2003,Bohren1998}. Moreover, by 
elongating or flattening the spherical NPs, one obtains spheroids which have 
two LSPRs corresponding to longitudinal and transverse polarization of 
light. For spheroids the plasmon resonance response depends solely on the 
aspect ratio, which is the ratio between the lengths of rotation axis and 
thickness of the particle \cite{Kelly2003,Bohren1998}. The 
prolate spheroids have the aspect ratio greater than 1, such that the 
longitudinal mode shifts to longer wavelength, whereas the transverse mode 
shifts to shorter wavelength. On the contrary, in oblate spheroids the 
longitudinal mode shifts to shorter wavelength, whereas the transverse mode 
shifts to longer wavelength. Thus by simply adjusting the aspect ratio, the 
LSPRs can be tuned at will over a broad range of wavelengths. 

The discrete-dipole approximation (DDA) \cite{Draine1994} the 
finite-difference time domain (FDTD) scheme \cite{Oubre2004}, and 
the boundary element method (BEM) in a full electromagnetic calculation 
\cite{GarciadeAbajo2002} have been successfully used in the 
calculation of the optical response of arbitrarily shaped NPs. These methods 
integrate the full Maxwell's equations but they are numerically extensive. 
Thus they cannot be directly used in the problem of designing plasmonic 
nanostructures. In addition, the above methods offer little inside about the 
formation, nature, and the behaviour of the LSPRs. As a more physical 
approach that works very well in the quasi-static limit, the hybridization 
model \cite{Prodan2003} has been proposed to solve some of the above 
issues regarding DDA, FDTD, and BEM methods. On the other hand, also in the 
quasi-static limit, the boundary integral equation (BIE) method \cite{Sandu2011,Fredkin2003} enables a direct relationship between 
the LSPRs and physical parameters of NPs like the shape (geometry) or the 
complex dielectric functions by relating the LSPRs to the eigenvalues and 
eigenfunctions of the operator associated with BIE. 

In the current work, the LSPR spectral modifications made through small but 
smooth shape variations from spheroids are studied by the BIE method of 
\onlinecite{Sandu2011}. Smooth deformations from spherical shape, prolate, and 
oblate spheroids are considered by keeping the same aspect ratio. The 
parameter that uniquely describes the shape variations can be related to the 
relative volume variation from spherical shape and from prolate and oblate 
spheroid, respectively. The influence of small shape changes on LSPRs has 
been treated in several works with emphasis on NP roughness \cite{Wang2006,Pecharroman2008,Trugler2011,Tinguely2011} or smooth shape variation \cite{Rodriguez2011}. All these 
papers monitor the spectral shift and, eventually, the inhomogeneous 
broadening of the main resonance due to shape variation. Recently a 
perturbative method has been developed in order to calculate the eigenvalue 
changes of LSPRs at small shape perturbations \cite{Grieser2009}. Despite 
many advantages like the use for designing plasmonic nanostructures with 
predetermined properties \cite{Ginzburg2011}, the method does not provide 
directly the weights of the LSPR eigenmodes and their modifications when 
small shape changes occur. On the other hand, in addition to the fact that 
it can be adapted to the recipes of \cite{Grieser2009}, the BIE method 
provides both the LSPR eigenvalues and their weights \cite{Sandu2011}. 
The present work shows that not only the eigenvalues change with the smooth 
modifications of the shape but also their weights change, sometimes in a 
drastic manner. There are also other goals of the present study. One other 
goal is related to the nanoparticle design and fabrication, which must be 
robust against the variation of nanoparticle shape. Thus during the 
fabrication processes like chemical synthesis or ``top-down'' approaches 
based on e-beam lithography variations of the process parameters are 
encountered. An example is the lift-off step in the top-down approach, where 
some precautions have to be made in order to fabricate systems with small 
features like dimers \cite{Acimovic2009} or oligomers \cite{Hentschel2010}. 
The present paper shows numerically that the eigenvalues and 
their weights have the largest variations when the applied field is along 
the smallest cross-sections. Hence additional care has to be considered in 
order to fabricate successful metallic NPs. Another reason comes from the 
simulations of experimental data, where one must find the most realistic 
model that fits the experimental setup. Thus in the simulation process some 
spectral features may be attributed only to the retardation and the shape 
variations are not considered at all. Generally, retardation moves the 
electrostatic resonances toward infrared \cite{Kelly2003,Bohren1998}. At the same time, some eigenmodes that are dark in the 
quasi-static limit become visible when the retardation is considered. In the 
present study it is shown that redshifting and the emergence of other higher 
order LSPRs may be obtained also by small shape changes. 

The paper has the following structure. The next section presents the BIE 
method and its accuracy. Then, the LSPR modifications due to smooth changes 
from various spheroidal shapes are analyzed in the section that follows the 
next section. The last section is dedicated to conclusions. 

\section{The Method and Its Accuracy}
In the electrostatic (quasistatic) approximation the optical behavior of the 
metallic NPs is described by the Laplace equation, whose solution may be 
obtained by the BIE method \cite{Sandu2011,Fredkin2003,Sandu2010}. The metallic NP delimited by surface $\Sigma $ is 
assumed to have a complex permittivity $\epsilon_i $ and is immersed in a medium of 
complex permittivity $\epsilon_o $. The incident electromagnetic field is 
represented as a uniform electric field ${\bf{E}}_0 $. The electric 
potential $\Phi $ associated with the total field ${\bf{E}}$ obeys the 
Laplace equation $\Delta \Phi \left( {\bf{x}} \right) = 0;\;{\bf{x}} \in \Re 
^3\backslash \Sigma $ with boundary conditions $\left. {\epsilon_0 \frac{\partial 
\Phi }{\partial {\bf{n}}}} \right|_ + = \left. {
\epsilon_1 \frac{\partial \Phi 
}{\partial {\bf{n}}}} \right|_ - ;\;{\bf{x}} \in \Sigma $ and $ - \nabla 
\Phi \left( {\bf{x}} \right) \to {\bf{E}}_0 ,\;\left| {\bf{x}} \right| \to 
\infty $. The total electric field is ${\bf{E}} =  - \nabla \Phi \left( 
{\bf{x}} \right)$, \textbf{n} is the normal to the surface $\Sigma 
$, and $\Re ^3$ is the 3D Euclidian space. The electric potential can be 
expressed in terms of the single-layer potential the surface charge $\mu 
_{E_0 } $ as

\begin{equation}
\label{eq1}
\Phi \left( x \right) = - {\bf{x}} \cdot {\bf{E_0}} + \frac{1}{4\pi }\int\limits_{{\bf{x}},{\bf{y}} \in \Sigma } 
{\frac{\mu _{E_0 } \left( {\bf{y}} \right)}{\left| {{\bf{x - y}}} \right|}dS_y } .
\end{equation}

The charge density $\mu _{E_0 } $ obeys the following integral equation

\begin{equation}
\label{eq2}
\frac{1}{2\lambda }\mu _{E_0 } \left( {\bf{x}} \right) - \hat {M}\left[ \mu \right] 
= {\bf{n}} \cdot {\bf{E}}_0 ,
\end{equation}

\noindent
where ${\bf{n}} \cdot {\bf{E}}_0 $ is the dot product of vectors in 3D, $\hat M$ is defined on 
surface $\Sigma $ as

\begin{equation}
\label{eq3}
\hat {M}\left[ \mu \right] = \frac{1}{4\pi }\int\limits_{{\bf{x}}, {\bf{y}} \in \Sigma } 
{\frac{\mu \left( {\bf{y}} \right) {\bf{n}} \left( {\bf{x}} \right) \cdot \left( {{\bf{x - y}}} 
\right)}{\left| {{\bf{x - y}}} \right|^3}d\Sigma _y } ,
\end{equation}

\noindent
and $\lambda = (\epsilon_i - \epsilon_o ) / (\epsilon_i + \epsilon_o )$ is a dielectric factor that 
depends on $\epsilon_o $ and $\epsilon_1 $. The operator $\hat M$ and its adjoint

\begin{equation}
\label{eq4}
\hat {M}^\dag \left[ \mu \right] = \frac{1}{4\pi }\int\limits_{{\bf{x}}, {\bf{y}} \in \Sigma 
} {\frac{\mu \left( {\bf{y}} \right) {\bf{n}} \left( {\bf{y}} \right) \cdot \left( {{\bf{x - y}}} 
\right)}{\left| {{\bf{x - y}}} \right|^3}d\Sigma _y } 
\end{equation}

\noindent
have the same discrete and real spectrum that is bounded between 1/2 and 
-1/2. In addition the number 1/2 is an eigenvalue irrespective of the NP 
shape. The charge density $\mu _{E_0 } $ is provided in terms of the 
eigenvalues and eigenfunctions of $\hat M$ and $\hat M^\dag $ as

\begin{equation}
\label{eq5}
\mu _{E_0 } = \sum\limits_k {\frac{1}{\textstyle{\frac{1} {2\lambda }} - \chi 
_k }} \left| {u_k } \right\rangle \left\langle {v_k } \right|\left. {{\bf{n}} \cdot 
{\bf{E}}_0 } \right\rangle .
\end{equation}

\noindent
where $\left| {u_k } \right\rangle ,\,\left| {v_k } \right\rangle $ are the 
eigenfunctions of $\hat M$ and $\hat M^\dag $, respectively, $\chi _k $ is the 
k$^{th}$ eigenvalue of $M$ and $M^\dag $, and $\left\langle {v_k } 
\right|\left. {{\bf{n}} \cdot {\bf{E}}_0 } \right\rangle $ is the scalar product of two 
square-integrable functions defined on $\Sigma $. The specific polarizabiliy 
of the NP that is the dipole moment generated by $\mu _{E_0 } $ divided by 
the NP volume $V$ has the form of a sum over all eigenvalues of $\hat M$ and $\hat M^\dag 
$

\begin{equation}
\label{eq6}
\alpha = \sum\limits_k {\frac{p_k }{\textstyle{\frac {1} {2\lambda }} - \chi 
_k }} .
\end{equation}

In Eq. (\ref{eq6}) $p_k = {\left\langle {{\bf{x}} \cdot {\bf{N}}} \right|\left. {u_k } 
\right\rangle \left\langle {v_k } \right|\left. {{\bf{n}} \cdot {\bf{N}}} \right\rangle }/ 
 V$ is the coupling weight of the 
$k^{th}$ eigenmode to the electromagnetic field and $N$ is the unit vector of 
the applied field given by $\vec {E}_0 = \vec{E}_0 \vec {N}$. The 
coupling weight ${p}_{k} $ is defined in terms of the eigenvectors 
of the operator $\hat M$ and its adjoint $\hat M^\dag $and it depends only on the 
shape of the NP, while the specific polarizability $\alpha $ depends also on 
the dielectric permittivities \cite{Sandu2010}. The eigenmodes with a 
non-zero $p_k $ couple with the light and they are called bright modes. On 
the other hand, many other eigenmodes do not couple with light and have 
vanishing $p_k $'s, thus they are called dark modes. Equation (\ref{eq6}) shows an 
eigenmode decomposition and allows an analytic expression for specific 
polarizability $\alpha $ when the dielectric permittivity of the metallic NP 
is described by a functional form like the Drude model $\epsilon\left( \omega 
\right) = \varepsilon_{m} - {\omega _{p}^2 }/
{\left( {\omega \left( {\omega + i\gamma } \right)} \right)}$. Here 
$\varepsilon _{m} $ is the interband contribution to the permittivity 
including $\varepsilon _\infty $, $\omega _p $ is the plasma frequency, 
$\gamma $ is the dumping constant, and $i = \sqrt { - 1} $. Thus for an 
embedding medium with a real and constant dielectric permittivity 
$\varepsilon _{d} $, the specific polarizability $\alpha $ in the 
Drude-like approximation takes the following form (Sandu et al. 2011)

\begin{equation}
\label{eq7}
\alpha (\omega) = 
\sum_k \frac{w_k (\varepsilon_m - \varepsilon_{d})}{\varepsilon_{eff\_k}}-\frac{w_k}{1/2-\chi_k} \frac{\varepsilon_{d}}{\varepsilon_{eff\_k}}
\frac{\tilde\omega_{pk}^2}{\omega(\omega + i\gamma) - \tilde\omega_{pk}^2},
\end{equation}

\noindent
where

\begin{equation}
\label{eq8}
\tilde {\omega }_{pk}^2 = \frac{\left( {1 / 2 - \chi _k } \right)\omega_p^2 
}{\varepsilon _{eff\_k} }
\end{equation}

\noindent
is the square of the localized plasmon resonance frequency in the limit of 
negligible $\gamma $ and $\varepsilon _{eff\_k} = \left( {1 / 2 + \chi _k } 
\right)\varepsilon_d + \left( {1 / 2 - \chi_k } \right)\varepsilon_m $ is 
an effective permittivity. The term $\left( {1 / 2 - \chi _k } \right)$ is 
the depolarization factor (DF) of the eigenmode \cite{Sandu2010,Sandu2011}. The specific polarizability $\alpha$ plays its role into the LSPR spectrum 
by its imaginary part that is proportional to the extinction spectrum of the incident light \cite{Bohren1998}.

To evaluate the polarizability $\alpha $ one needs to calculate the 
eigenvalues and the eigenfunctions of $\hat M$ and $\hat M^\dag $ using a basis set of 
functions that are defined on surface $\Sigma $. If the surface $\Sigma $ 
can be parameterized by $\{x = g(z)\cos \varphi ,y = g(z)\sin \varphi ,z\}$, 
where $g(z)$ is a smooth but otherwise arbitrary function, the variables 
that determine the surface are the coordinate $z$ and the angle $\varphi $, 
respectively. This parameterization allows the surface $\Sigma $ to be 
smoothly mapped onto unit sphere. Here $g\left( z \right)$ is sufficiently 
smooth defined on $z \in \left[ {z_{min} ,z_{max} } \right]$, such that 
$g\left( {z_{max} } \right) = g\left( {z_{min} } \right) = 0$, and 
${g}'\left( {z_{max} } \right) = {g}'\left( {z_{min} } \right) \to 
\infty $. By a linear transformation of the form $z = \frac{1}{2}\left( 
{\left( {z_{max} - z_{min} } \right)z_1 + z_{max} + z_{min} } 
\right)$, the new variable $z_1 $ is restricted to $\left[ { - 1,1} 
\right]$, such that the one-to-one mapping onto a sphere becomes now 
apparent. Thus. through the above mapping, the basis of spherical harmonics 
$Y_{lm} \left( {\theta ,\varphi } \right)$ defined on the unit sphere 
generates the basis of functions on $\Sigma $. All the other details about 
numerical procedure can be found in \onlinecite{Sandu2010}. 

\begin{figure}
  \includegraphics{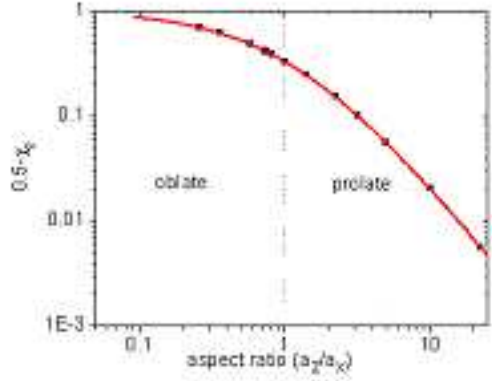}
\caption{The comparison between the numerical results (the black dots) and 
the analytical results \cite{Bohren1998} (the red full curve) of the 
DF of spheroids with various aspect ratios.}
\label{fig:1}      
\end{figure}

The numerical accuracy of the method will be verified below for spheroids of 
various aspect ratio values. For ellipsoids and, in particular, for 
spheroids, there are three equally weighted bright modes determined by the 
electric field polarizations that are parallel to each axis. These bright 
modes are, in fact, the dipole eigenmodes, of which two of them are 
degenerate for spheroids. In addition to that the DFs of spheroids can be 
expressed analytically as functions of the aspect ratio defined by the term 
${a_z }/{a_x }$ (spheroids being considered to be the 
ellipsoids of semiaxes $a_x $, $a_y $, $a_z $, such that $a_x = a_y )$ 
\cite{Bohren1998}. Thus, for spheroids, the following relations 
must be fulfilled: $p_x = p_y = p_z = 1$, $\chi _x = \chi _y $, and the$\chi 
_x + \chi _y + \chi _z = 1/2$ (Bohren and Huffman 1998; Sandu et al. 2010). 
In Fig. \ref{fig:1} the calculated DFs are compared to the analytical expressions 
provided in \onlinecite{Bohren1998}. A relative accuracy of at least 10$^{ 
- 5}$ is obtained by using a basis of only 25 functions and a numerical 
quadrature of 96 Gauss points \cite{Sandu2010}. Axially symmetric NPs 
permit much faster and more stable and accurate solutions due to the fact 
that numerical integration is performed only in 2D \cite{Geshev2004,Geshev2006,Pedersen2011}. In fact, axial symmetry allowed 
us to obtain stable bright eigenmodes whose eigenvalues are close to 1/2, 
which is the largest eigenvalue of the system and is also dark \cite{Sandu2010,Sandu2011}. 

\section{The Effect of Shape Variation}
\subsection{The shape}

Since the spheroids have an analytical form for their DFs, the variation of 
the LSPR response with respect to the aspect ratio change can be easily 
assessed. In the present work smooth shape changes that keep the same aspect 
ratio will be considered. Thus the NP shapes that will be taken into account 
below have the form

\begin{equation}
\label{eq9}
g\left( z \right) = C\left( {z,b} \right)\sqrt {1 - \left( {{\frac{z} 
{a}}} \right)^2} .
\end{equation}

The parameters $a$ and $b$, and the smooth function $C\left( {z,b} \right)$ 
establish the shape of an individual particle. Without $C\left( {z,b} 
\right)$, Eq. (\ref{eq9}) describes a spheroid with an aspect ratio given by $a$ and 
$z_{max } = - z_{min } = a$. Since the results of the quasistatic 
approximation are scale-invariant, just one parameter suffices to define a 
spheroid, which is usually described by two parameters (i. e., its distinct 
semiaxes). The function $C\left( {z,b} \right)$ quantifies the smooth 
deviation from the spheroidal shape and is chosen to be peaked in the 
vicinity of $b = 0$, such that the spheroid is determined by $b = 0$. Thus 
the following form of $C\left( {z,b} \right)$ is adopted

\begin{equation}
\label{eq10}
C\left( {z,b} \right) = \frac{2}{1 + \left( {1 - bz^2} \right)^2}.
\end{equation}

Equations (\ref{eq9}) and (\ref{eq10}) cover a large class of shapes including shapes 
similar to nanodisks and nanorods. For example, a nanorod with an aspect 
ratio of 4 can be generated by taking b=0.04 and a=4 and a nanodisk with an 
aspect ratio of 1/2 can be made by having b=2 and a=0.5. Moreover, instead 
of parameter $b$, one can choose a more intuitive parameter, the relative 
volume variation from a spheroid, as a shape parameter. Thus relative volume 
variations ${\Delta V}/ V$ are negative/positive for negative/positive 
$b$. 

In the calculations of the NP polarizability only the bright modes 
contribute to LSPR spectrum. Smooth deformations from spheroidal shape given 
by Eq. (\ref{eq9}) will determine also higher order eigenmodes to become bright. For 
some of these bright modes, however, their weights may not be large enough 
to be distinguishable from the background or from other modes due to plasmon 
ovelapping. If $\gamma $is much less than the distance between two 
consecutive plasmon frequencies $\tilde {\omega }_{pk} $, then the 
overlapping is not encountered and the following criterion

\begin{equation}
\label{eq11}
\frac{p_k }{1 / 2 - \chi _k }\frac{\varepsilon _d }{\varepsilon _{eff\_k} 
}\frac{\tilde {\omega }_{pk} }{\gamma } \ge 1
\end{equation}

\noindent
may be used to determine if a bright mode shows up into the LSPR spectrum. 
Eq. (\ref{eq11}) can be used as a guiding criterion for the relevance of an 
eigenmode to the LSPR spectrum. It says that the height of the LSPR should 
be larger than a baseline of magnitude one. Thus for a ratio ${\tilde 
{\omega }_{pk} }/ \gamma \approx 10$, which is perfectly 
normal, ${\varepsilon _d } / \approx 1$ (also within a typical range), and $1 / {\left( {1 / 2 - \chi _k } \right)} > 1$, the 
weights fulfilling the condition $p_k > 0.1$ would allow (\ref{eq11}) to be true and 
the bright eigenmodes to be observed in the spectrum.

\subsection{Changes due to the variations from spherical shape}

\begin{figure}
  \includegraphics{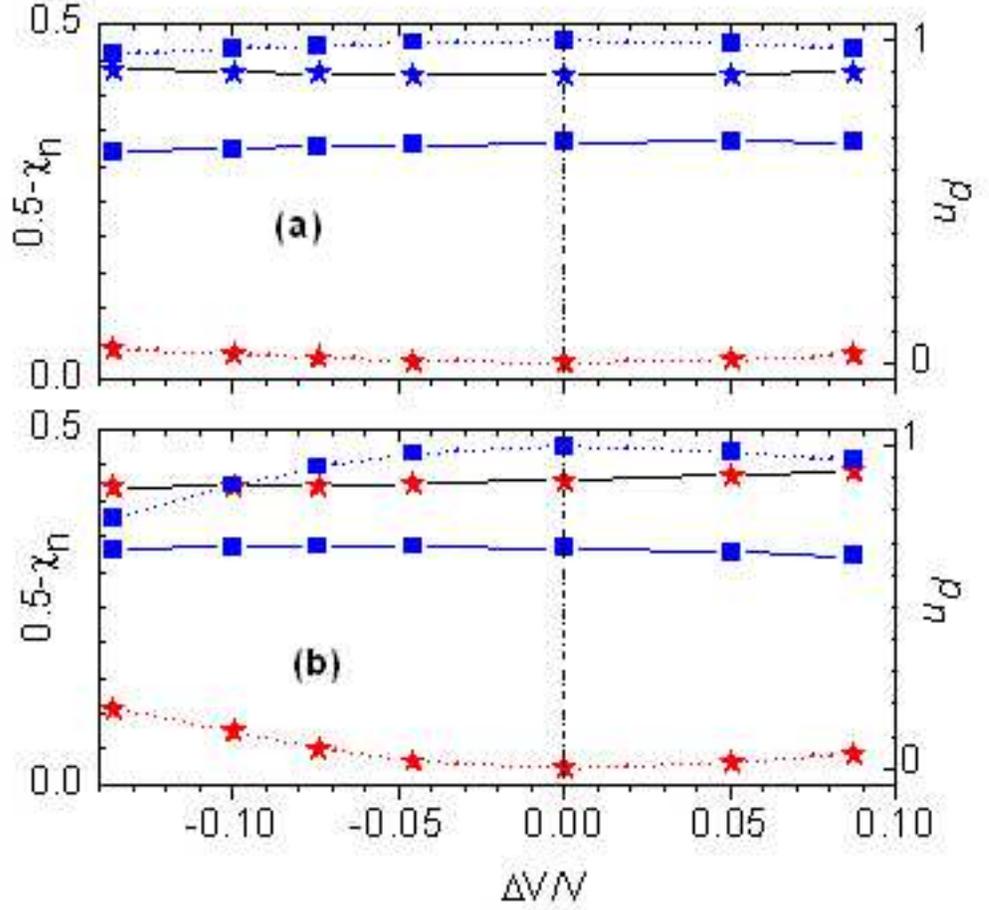}
\caption{ Changes of the two most relevant DFs (left side and continuous 
lines) and of their weights (right side and dotted lines) with respect to 
shape variation from sphere. The shape variation is accounted by the volume 
variation $\Delta V/V$. The first (dipole)/the second (octopole) eigenmode is denoted by 
the square/star symbol. The field polarization is (a) parallel or (b) 
perpendicular to the rotation axis.}
\label{fig:2}       
\end{figure}

\begin{figure}
  \includegraphics{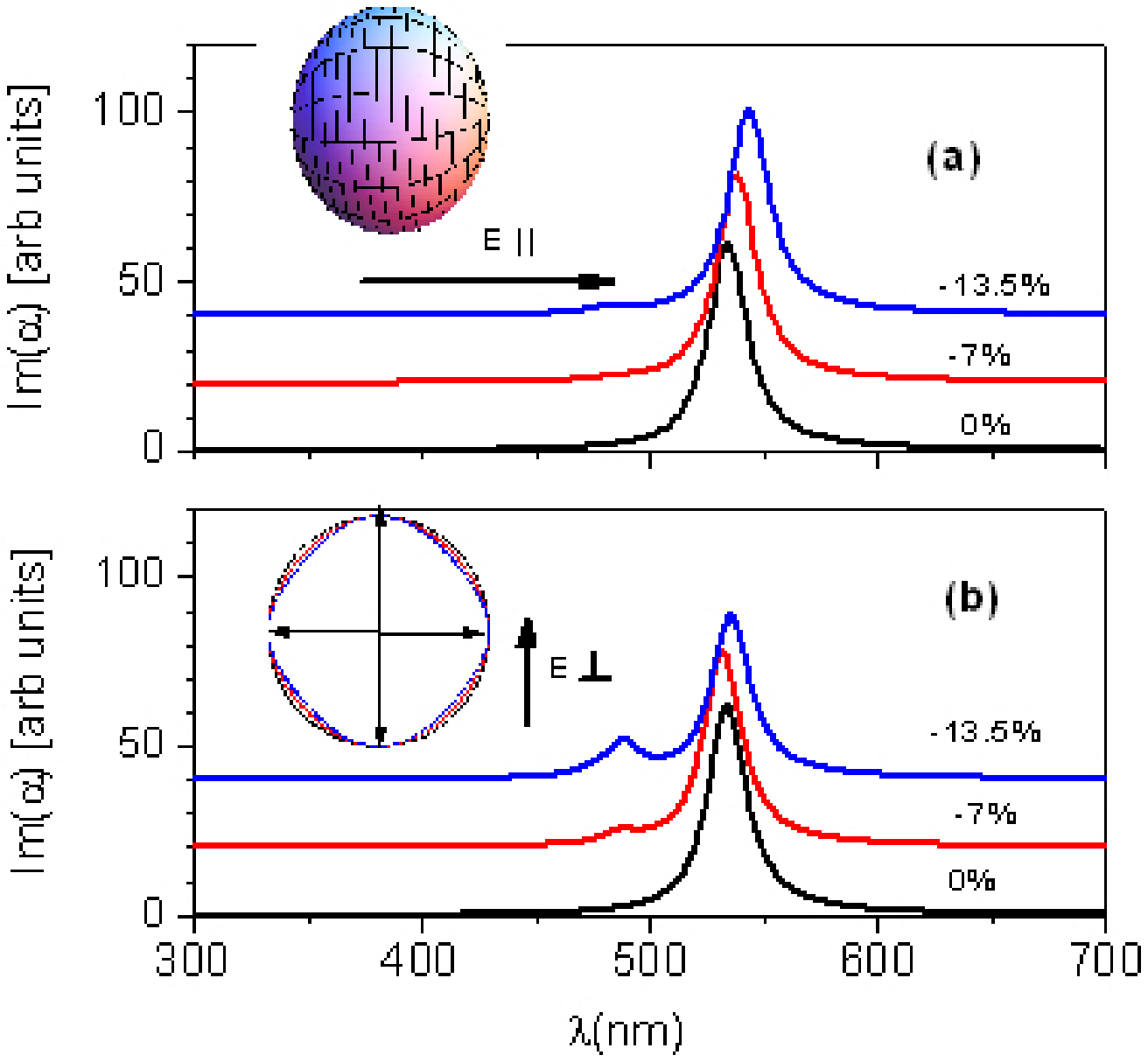}
\caption{ The LSPR spectra for three different $\Delta V/V$: -0{\%}, -7{\%}, and 
-13.5{\%}. For a better visualization the spectra are moved upwards 
accordingly. The field polarization is (a) parallel or (b) perpendicular to 
the rotation axis. The inset of (a) depicts a sphere and the field 
polarizations. The inset of (b) represents the cross-section variations 
corresponding to shape variations.}
\label{fig:3}       
\end{figure}

Throughout this work I consider small shape variations of gold NPs immersed 
in water ($n_{water}$=1.33). The Drude parameters of gold are those that are 
usually used in the literature. Thus, in energy, the plasma frequency 
$\omega _p $ is about 9 eV and $\gamma $ is about 0.1 eV \cite{Sandu2011}. Also the interband $\varepsilon_{m} $ is adjusted to have the 
LSPR for a nanosphere at 530 nm, i.e., $\varepsilon_{m} = 10.2$. Fig. 
\ref{fig:2} illustrates the variations of the two most relevant eigenvalues (or 
equivalently the associated DFs) and their weights with respect to shape 
variation from a nanosphere and field polarization. An eigenmode is 
considered relevant if its weight may exceed 1{\%}. The relative volume 
${\Delta V}/ V$ varies with respect to spherical shape from 
-14{\%} to 10{\%}. Fig. \ref{fig:3} presents the variation of LSPR spectra with 
respect to both shape variation and field polarization. The shape variations 
are hardly detected by simple vizualization as one can see from the inset of 
Fig \ref{fig:3}b, which shows the variations in cross-section for volume variations up 
to 14{\%}. 

For parallel polarization, the first relevant eigenmode represents the 
dipole response and has a monotonic ( i. e. increasing) behavior of its DF 
with respect to volume variation. Therefore, according to Eq. 7 a smaller DF 
means a redshifted LSPR as it is observed from Fig \ref{fig:2}a and \ref{fig:3}a at negative 
volume variations. The NP is invariant with respect to coordinate change $z 
\leftrightarrow - z$, thus the quadrupole eigenmode is not allowed to be 
bright, therefore the second most relevant eigenmode is the octopolar 
eigenmode. In contrast to the dipolar one, the octopolar eigenmode exhibits 
a non-monotonic variation of the DF with a minimum reached when the particle 
is spherical ($\Delta V/V = 0$\%). No major variation of its weight (i. e. no more than 
10{\%}), however, occurs with no noticeable change in the LSPR spectrum. 
Thus, from Fig \ref{fig:3}a and according to the first sum rule $\sum\limits_k {p_k } 
= 1$ (Sandu et al. 2010), the weight variation of the dipole eigenvalue is 
also less than 10{\%}.

There is a different picture for transverse polarization (Fig. \ref{fig:2}b and Fig 
\ref{fig:3}b). The DF of the dipole eigenvalue is non-monotonic with a maximum around 
$\Delta V/V = -10 $ \% and its weight varies by -20{\%} for a volume variation of 
-13.6{\%}. Moreover, for positive $\Delta V/V$, the LSPR of the dipole eigenmode 
redshifts as a result of the DF drop. The large reduction of the dipolar 
weight is reflected in large increase of the octopolar weight, which can be 
also seen in the LSPR spectrum plotted in Fig \ref{fig:3}b. At the same time the DF of 
the octopolar eigenmode increases with $\Delta V/V$.

\subsection{Changes due to the variation from prolate shape}

\begin{figure}
  \includegraphics{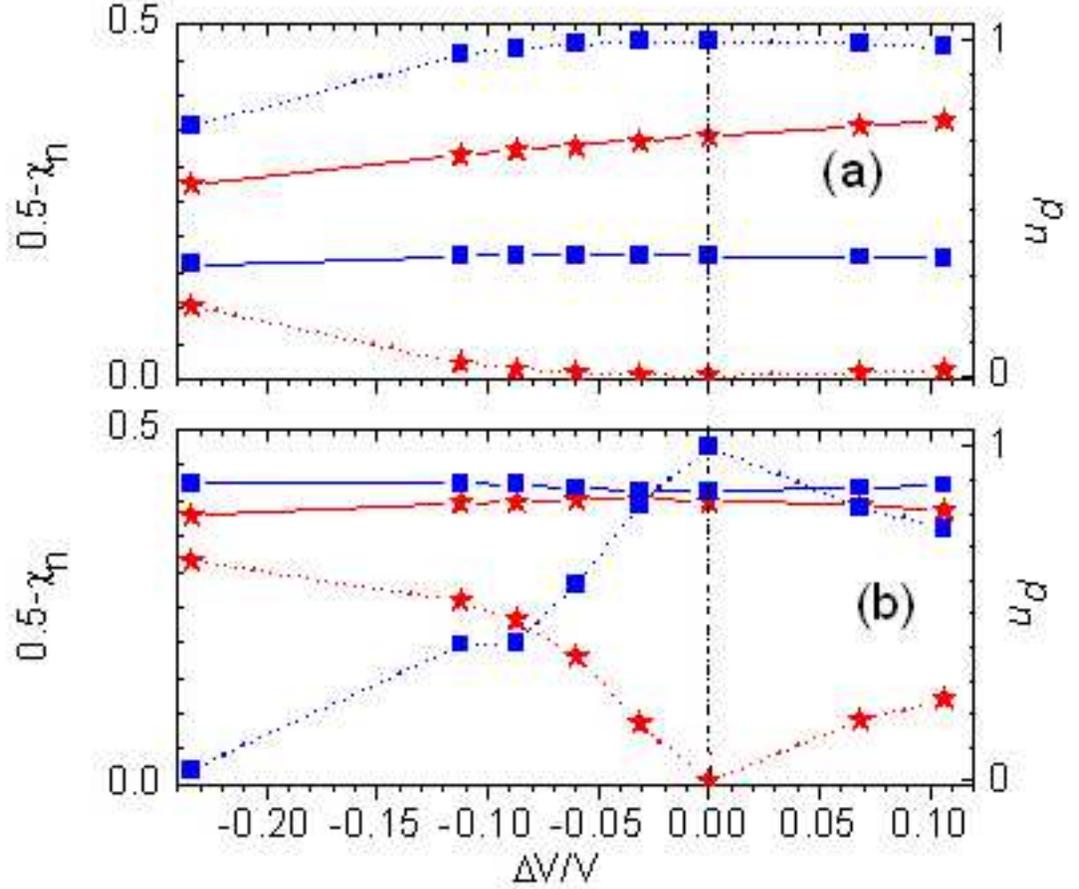}
\caption{Changes of the two most relevant DFs (left side and continuous 
lines) and of their weights (right side and dotted lines) with respect to 
shape variation from prolate spheroidal shape. The shape variation is 
accounted by $\Delta V/V$. The first (dipolar)/second (octopolar) eigenmode is denoted 
by the square/star symbol. The field polarization is (a) parallel or (b) 
perpendicular to the rotation axis.}
\label{fig:4}       
\end{figure}

\begin{figure}
  \includegraphics{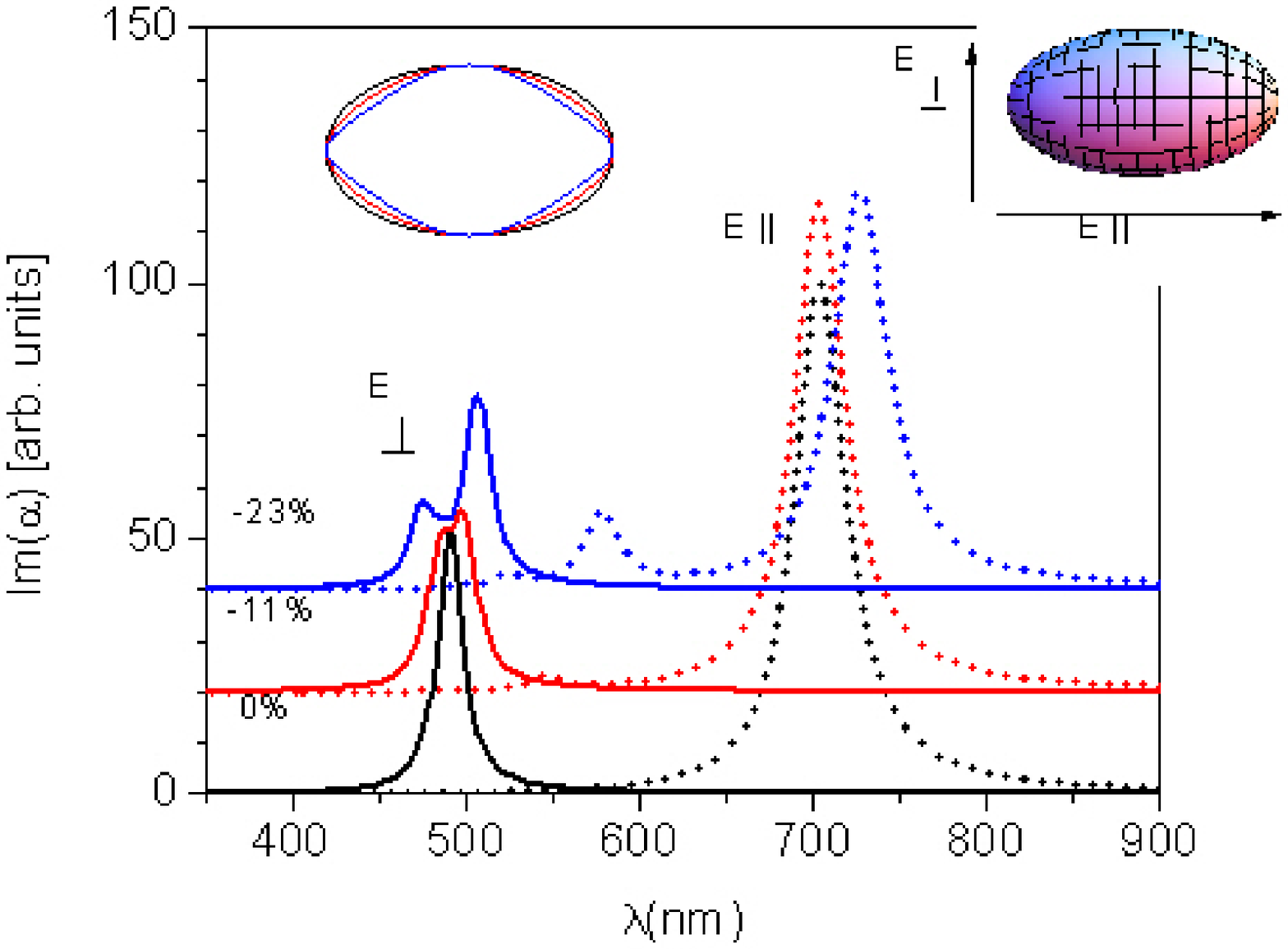}
\caption{The LSPR spectra for three different $\Delta V/V$: 0{\%}, -11{\%}, and -23{\%}. 
For a better visualization the spectra are moved upwards accordingly. The 
field polarization is parallel (dotted lines and the LSPRs at longer 
wavelengths) or perpendicular (solid lines and the LSPRs at shorter 
wavelenghts) to the rotation axis. The insets depict a prolate spheroid and 
the field polarizations (right side) and the axial cross-section variations 
corresponding to shape variations (left side).}
\label{fig:5}      
\end{figure}

The changes of the electrostatic resonances due to smooth variation from a 
prolate shape are shown in Fig. \ref{fig:4}, where only the two most relevant 
eigenvalues and their weights are presented. The aspect ratio is 2. For both 
field polarizations, ${\Delta V}/ V$ varies over a wider range from -24{\%} 
to 11{\%}. The behavior of LSPR spectra with respect to shape variations can 
be seen in Fig. \ref{fig:5}. The shape variations that are defined by volume 
variations can be vizualized in the left inset of Fig \ref{fig:5}. Like in the 
previous case, the quadrupole eigenmode is dark, regardless of shape change. 
When the polarization is longitudinal, the DFs of the most relevant 
eigenmodes (dipolar and octopolar) vary smoothly over the entire range of 
volume variation considered here. The corresponding LSPRs can be noticed at 
longer wavelengths in Fig. \ref{fig:5}. While the DF of the octopole eigenmode 
increases with ${\Delta V}/V$, the DF of the dipole eigenmode 
reaches its maximum at ${\Delta V}/ V = - 11\% $. The LSPRs are sensitive 
to shape variations, such that at ${\Delta V}/ V = - 11\% $, an 
octopolar bump can be seen in the spectrum. The bump transforms into a 
distinct LSPR at ${\Delta V}/ V = - 23\% $. 

\begin{figure*}
\centerline{\includegraphics[width=7.5in,height=2.36in]{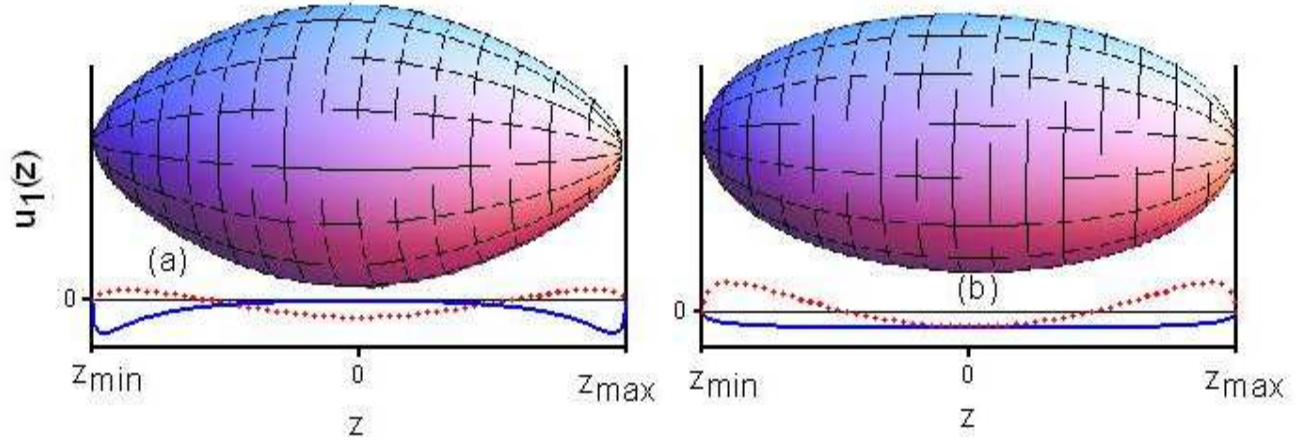}}
\caption{The z-dependent part u$_{1}$(z) of the dipole (solid line) and 
octopole (dotted line) eigenfunctions for (a) football-like and (b) prolate 
spheroidal NPs. For transverse polarization the full eigenfunctions have an 
extra factor $\sin \phi $. The shape of these NPs are depicted in the insets.}
\label{fig:6}       
\end{figure*}

The behavior is however quite different for transverse polarization, where 
the cross-section is smaller. Dramatic changes in the weights of the 
relevant eigenmodes are found although the eigenvalues and their DFs vary 
smoothly. The appearance of the octopolar eigenmode in the LSPR spectrum 
occurs for a volume variation as small as $\pm 5\% $. The DF of the dipole 
eigenmode is non-monotonic with a minimum at $\Delta V/V = 0$ and a maximum around 
$\Delta V/V = -11 \%$. An interesting behavior occurs around $\Delta V/V = -10 \% $, where the shape 
looks rather like an American football (Fig. \ref{fig:6}a) and the weight of the 
octopolar eigenmode surpasses the weight of the dipolar eigenmode. The 
weight of the dipole eigenmode varies from 100{\%} at $\Delta V/V = 0$ to a value of 
around 40{\%} for a volume variation of -11{\%} (Fig \ref{fig:4}b). Moreover, for 
small positive and negative $\Delta V/V$, the LSPR of the dipole eigenmode blueshifts as 
a result of the DF increase. The large reduction of the dipolar weight is 
reflected in large increase of the octopolar weight, which can be also seen 
in the LSPR spectrum plotted in Fig \ref{fig:5}b. At the same time the DF of the 
octopolar eigenmode reaches its maximum at $\Delta V/V = -3 \%$. 

To understand why the octopole weight exceeds the dipole weight it is better 
to inspect the dipole and the octopole eigenfunctions of $\hat M$ for both 
shapes: the spheroidal prolate shape and the football-like shape obtained by 
a volume variation of -11{\%}. These eigenfunctions are plotted in Fig. \ref{fig:6}. 
Due to the axial symmetry, for a transverse polarization, the eigenvectors 
have the following expression $u\left( {z,\varphi } \right) = u_1 \left( z 
\right) \sin \varphi $, when the field is along the y-axis. The 
functions $u_1 \left( z \right)$ of both the dipole and the octopole 
eigenmodes are plotted in Fig. \ref{fig:5}. Thus the dipole eigenmode represents a 
large dipole ($ \uparrow )$ concentrated in the middle of the NP in the case 
of spheroidal shape (Fig. \ref{fig:6}b) and splits into two parallel dipoles ($ 
\uparrow  \uparrow )$ towards the extremities for a football-like shape 
(Fig. \ref{fig:6}a). These two parallel dipoles are much smaller than the spheroidal 
dipole because the extremities are much tighter than the middle of the NP. 
It is worth noticing that the dipolar character of the eigenmode is provided 
by the factor $\sin \varphi $ of the eigenfunction. At the same time, the 
octopole eigenmode has the dipole configuration $ \downarrow  \uparrow  
\downarrow $ but the total dipole moment vanishes for the spheroidal shape 
and it is different from 0 for the football shape. Also the dipolar 
contribution of the octopole eigenmode exceeds the contribution of the 
dipole eigenmode because the football-like NP is much thicker in the middle 
than toward the extremities. 

\subsection{Changes due to the variation from oblate shape}

\begin{figure*}
  \includegraphics{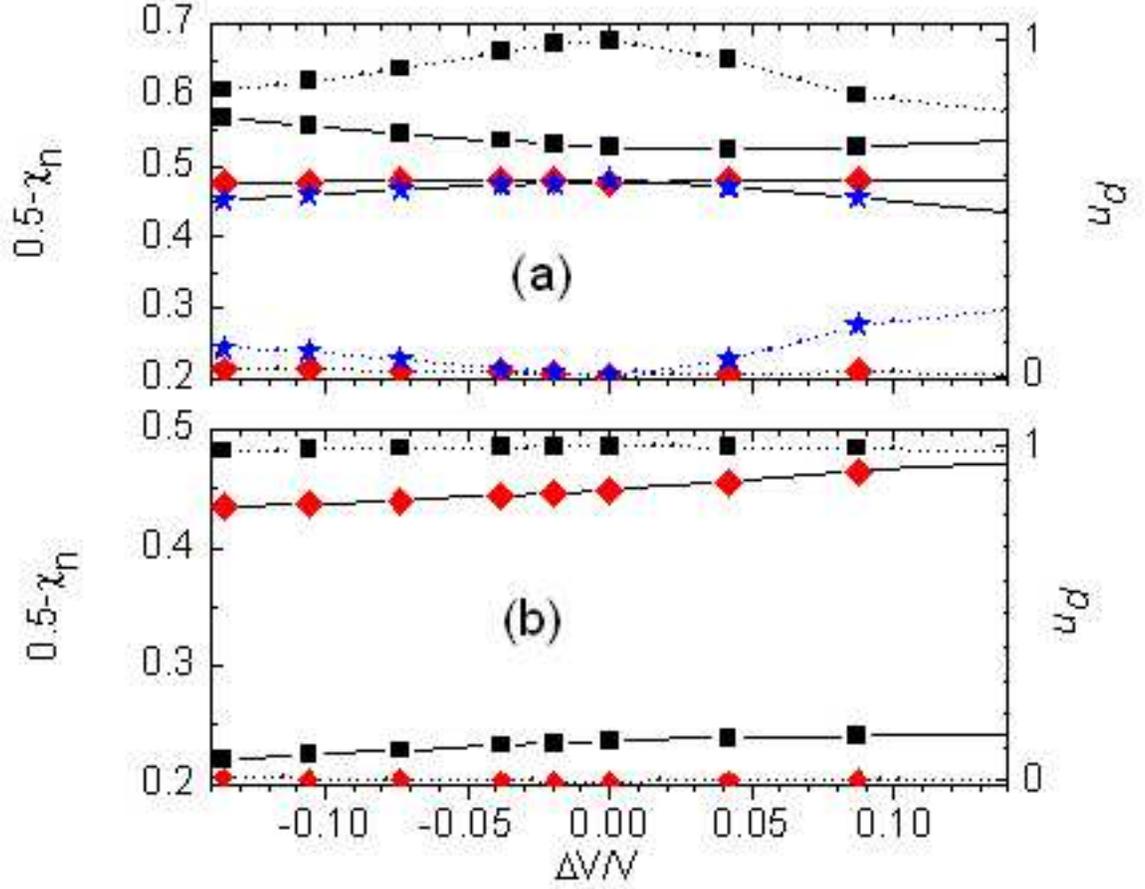}
\caption{ Changes of the relevant DFs (left side and continuous lines) and of 
their weights (right side and dotted lines) with respect to shape variation 
from oblate spheroidal shape. The first eigenmode (dipolar) is denoted by 
square symbol. The other eigenmodes are denoted by star and diamond symbols. 
The field polarization is (a) parallel or (b) perpendicular to the rotation 
axis.}
\label{fig:7}       
\end{figure*}

\begin{figure*}
  \includegraphics{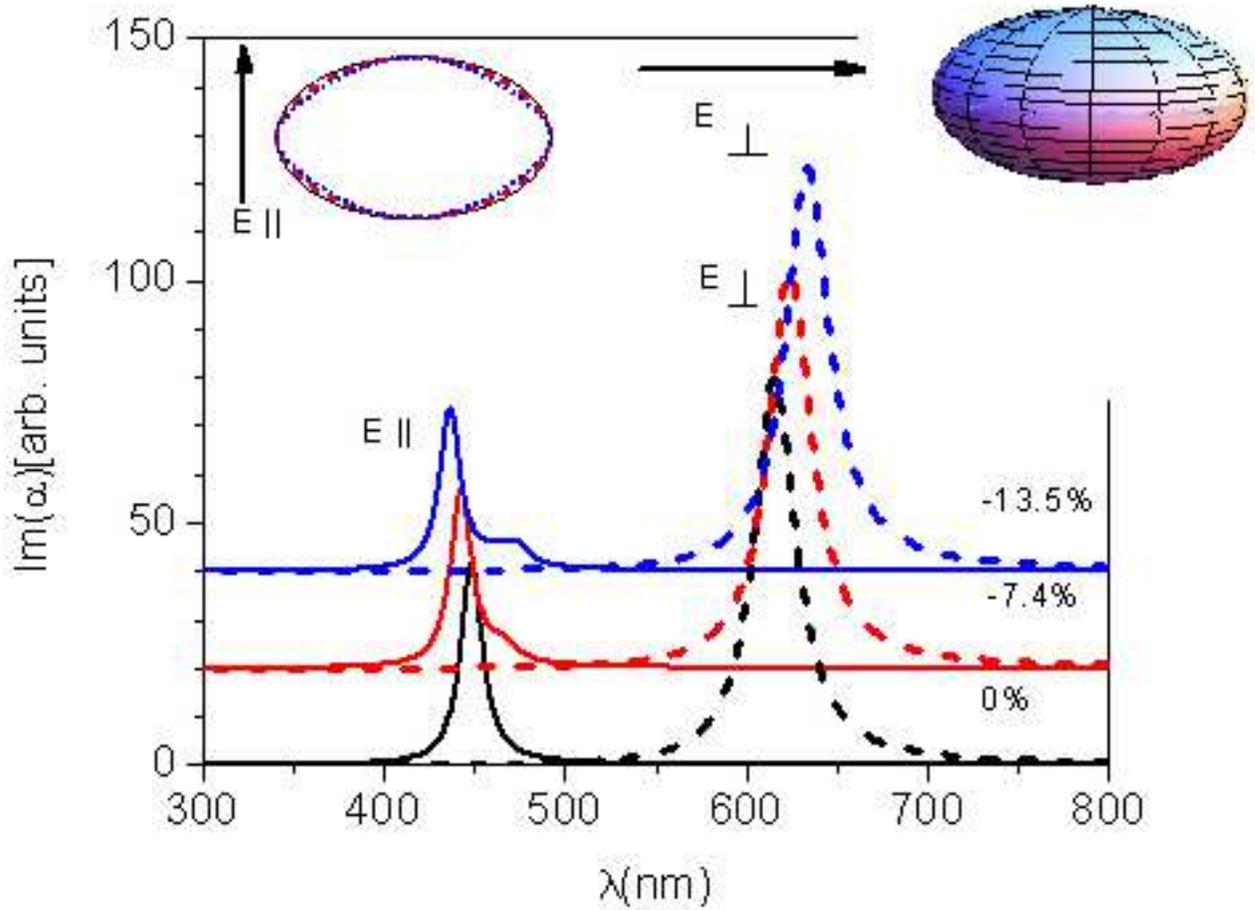}
\caption{The LSPR spectra for three different $\Delta V/V$: 0{\%}, -7.4{\%}, and 
-13.5{\%}. For a better visualization the spectra are moved upwards 
accordingly. The field polarization is parallel (full line and the LSPRs at 
shorter wavelengths) or perpendicular (dotted line and the LSPRs at longer 
wavelenghts) to the rotation axis. The inset depicts an oblate spheroid 
(right side) and the axial cross-section variations corresponding to volume 
variations (left side).}
\label{fig:8}      
\end{figure*}

The variations of the DFs and of their weights for the most relevant 
eigenvalues with respect to shape variation of an oblate spheroid and field 
polarization are presented in Fig. \ref{fig:7}. The aspect ratio is taken to be 1/2. 
The axial polarization exhibits three relevant eigenmodes, while the 
transverse polarization has just two relevant eigenmodes. The variation of 
the relative volume ${\Delta V}/ V$ spans between -14{\%} and 14{\%}. Fig. \ref{fig:8} 
shows the variation of LSPR spectra with respect to both shape variation 
and field polarization. 

The electric field parallel with the rotation axis is considered first. In 
this case the LSPRs are found at shorter wavelenghts in Fig. \ref{fig:8}. The first 
relevant eigenmode represents also the dipole response and its corresponding 
DF has a non-monotonic behavior by having the minimum at ${\Delta V}/V = 4\% $. Thus there is a blueshift of the 
corresponding LSPR for negative $\Delta V/V$ (see Fig. \ref{fig:8}). The next most relevant is the 
octopole eigenmode which can be seen as a bump for a volume variation of 
-7.4{\%} and a quite distinct new resonance at ${\Delta V}/ V = - 13.5\% 
$. The weight of the third relevant eigenmode does not exceed 2{\%} over 
entire range of $\Delta V/V$, therefore this eigenmode is not seen in the LSPR spectrum. 
The variations of DFs of the octopole and the third relevant eigenmode are 
non-monotonic. The octopole eigenmode has a minimum at ${\Delta V}/V = 0$ and a maximum 
around ${\Delta V}/ V = - 7.5\% $, while the third relevant eigenmode 
has a maximum at ${\Delta V}/ V = 0$. For transverse polarization the 
modifications of the DFs of the two most relevant eigenmodes (the dipole and 
the octopole) increase with $\Delta V/V$ such that for negative/positive $\Delta V/V$ there is a 
redshift/blueshift of the dipole LSPR. The calculations also show that the 
octopole eigenmode does not show up in the LSPR spectrum. 

\section{Conclusions}

In the quasistatic approximation I use a BIE method to analyze the changes 
of LSPRs produced by small and smooth variations of the shape of metallic 
NPs. The LSPRs are determined by the eigenvalues and their coupling weights 
that come from the operator associated to the BIE method. First, the 
numerical implementation of BIE is verified against well known results. Thus 
for axially symmetric structures the method has an excellent accuracy. Also 
a compact formula of the NP polarizability is used to elaborate a criterion 
for the structural changes of the spectrum with respect to shape variation. 
Such shape variations are inherent to NP fabrication process. Shape 
variations induce changes in both the eigenvalues and their coupling weights 
to the electromagnetic field. However, the coupling weights appear to be 
more affected by shape changes, such that they manifest as structural 
modifications of the LSPR spectrum. Full numerical calculations show that 
structural modifications of the optical response occur for shape changes 
corresponding to volume variations greater than 12{\%}. This quantitative 
result can be directly related to the criterion provided by Eq. (\ref{eq11}), where 
a variation of $p_k > 0.1$ would generate structural changes in LSPR 
spectrum. A 0.1 variation of $p_k $ means a 10{\%} volume variation via the 
definition of $p_k $, which is inverse proportional to volume $V$. Therefore 
the NP volume variation contributes significantly to the structural changes 
of the LSPR spectrum which is agreement to other results regarding the 
averaging effect of the roughness on the LSPRs of metallic NPs \cite{Trugler2011}. 
The largest optical changes are encountered for light 
polarization that is parallel to the smallest cross-sections of NPs. This 
aspect is important because more care must be considered for the design and 
the fabrication of the smallest features of the NP systems.
\begin{acknowledgments}
This work has been supported by the Sectorial Operational Programme Human Resources Development, 
financed from the European Social Fund and by the Romanian Government under the contract number POSDRU/89/1.5/S/63700. 
\end{acknowledgments}


\end{document}